\documentclass[3p,a4paper,times]{elsarticle}
\usepackage{graphicx,xcolor,caption,subcaption,float,lipsum,amsmath,amssymb,amsfonts,amsthm,amsbsy,amstext,stmaryrd}
\usepackage{lineno}
\usepackage{siunitx}
\usepackage{xfrac}
\usepackage{booktabs}
\usepackage[english]{babel}
\usepackage{multicol}

\biboptions{square,comma,sort&compress}

\begin{document}

\begin{frontmatter}
\title{Ground state of the U$_2$Mo compound: Physical properties of the $\Omega$-phase.}

\author[cab1]{E. L. ~Losada}
\author[cab2]{J.E.~Garc\'es$^\ast$}  
\address[cab1]{SIM$^3$, Centro At\'omico Bariloche, Comisi\'on Nacional de Energ\'ia At\'omica, Argentina}
\address[cab2]{Gerencia de Investigaci\'on y Aplicaciones Nucleares, Comisi\'on Nacional de Energ\'ia At\'omica, Argentina}

\begin{abstract}
Using \textit{ab initio} calculations, unexpected structural instability was recently found in the ground state of the U$_2$Mo compound. Instead of the unstable $I4/mmm$ structure, in this work the $P6/mmm$ ($\# 191$) space group, usually called $\Omega$-phase, is proposed as the fundamental state. Electronic and elastic properties are studied in this work in order to characterize the physical properties of the new ground state. 
The stability of the $\Omega$-phase is studied by means of its elastic constants calculation and phonon dispersion spectrum. Analysis of isotropic indices shows that the new phase is a ductile material with a minimal degree of anisotropy, suggesting that U$_2$Mo in the $P6/mmm$ structure is an elastic isotropic material. Analysis of  charge density, density of electronic states (DOS) and the character of the bands revealed a high level of hybridization between $d$-molybdenum electronic states and $d$- and $f$-uranium ones.

\end{abstract}

\end{frontmatter}

\long\def\symbolfootnote[#1]#2{\begingroup%
\def\thefootnote{\fnsymbol{footnote}}\footnote[#1]{#2}\endgroup}
\long\def\symbolfootnotemark[#1]{\begingroup%
\def\thefootnote{\fnsymbol{footnote}}\footnotemark[#1]\endgroup}
\long\def\symbolfootnotetext[#1]#2{\begingroup%
\def\thefootnote{\fnsymbol{footnote}}\footnotetext[#1]{#2}\endgroup}
\symbolfootnotetext[1]{Corresponding author email: garces@cab.cnea.gov.ar}

\section{INTRODUCTION}\label{sec.introd}
\label{Intro}

Renewed interest has recently been shown in the U-Mo system. The main reasons are related to its relevance as a potential nuclear fuel to be used in Material Testing Reactor (MTR) \cite{Snelgrove1997} and $ GenIV $ power reactors \cite{Kim2013}. However, it is also due to the unusual physical properties found in the ground state of this system \cite{Losada2015}. 
The unique stable compound observed in the phase diagram at finite temperature is U$_2$Mo. The ground state of this compound was previously assumed to be the C11b structure (MoSi2 prototype, $I4/mmm$ space group), since there was no information to contradict this hypothesis. Detailed theoretical calculations also found one compound in the ground state of the U-Mo system \cite{Landa11}. 
Nevertheless, it was shown in Refs.  \cite{Wang2014,Liu2015} that the assumed C11b structure is unstable under the $D_6$ deformation associated with the $C_{66}$ elastic constant. It was proposed in Ref. \cite{Wang2014}, using the USPEX code, that the ground state of the U$_2$Mo compound has the $Pmmn$ structure.

\begin{figure}[!h]
\centering
\includegraphics[width=8cm]{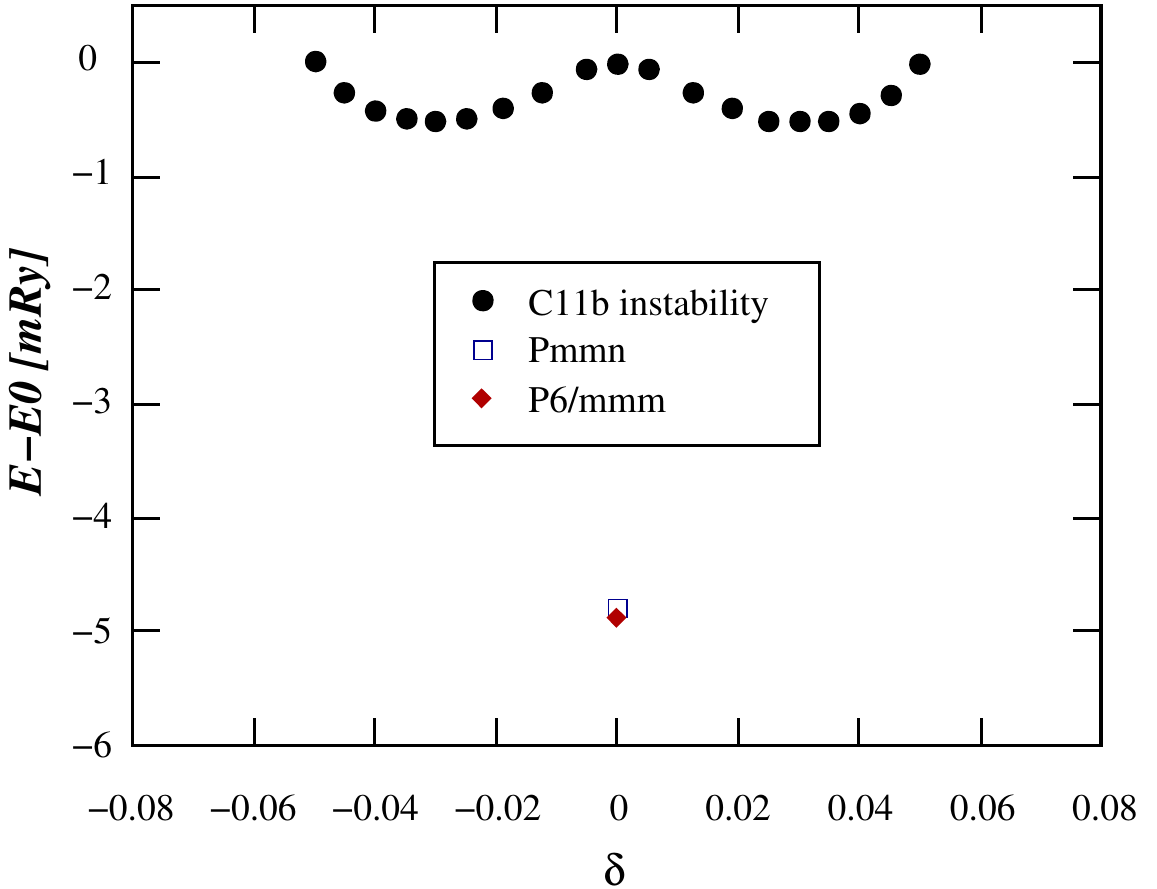}
\captionsetup{justification=centering}
\caption{Comparison between energies of different structures of the U$_2$Mo compound. Filled circles points correspond to the total energy difference between the undistorted $I4/mmm$ structure ($\delta =0$) and the structures obtained by applying small variations to the $D_6$ distortion parameter, as implemented in Ref. \citep{Losada2015}. $E_0$ is the total energy corresponding to the $I4/mmm$ structure. The data points labelled with an empty square and a filled diamond correspond to the energy difference between $I4/mmm$ and the $Pmmn$ and $P6/mmm$ structures, respectively.
 \label{fig:compMin}}
\end{figure}

The same result regarding the above-mentioned instability was obtained in Ref. \citep{Losada2015} through  analysis of the elastic constants. The instability associated with the $D_6$ distortion of the C11b structure is shown in Fig. \ref{fig:compMin}. 
The results published in Ref.\cite{Losada2015} showed that the structure of the ground state of the U$_2$Mo compound is other than the assumed C11b and the $Pmmn$ structures. Taking into account more isotropic atomic arrangements and computing the total energy using the WIEN2k code \cite{Wien2k}, a more symmetric structure was found, which is energetically more stable than the $Pmmn$ one. The P6 space group (\# 168) was proposed as the ground state of the U$_2$Mo compound, as it is 0.1 mRy more stable than the $Pmmn$ structure, the total energy of both structures being computed in the primitive unit cell. 
The root causes of the structural instability were also explained in Ref. \cite{Losada2015}. It was found that the distorted structure is stabilized due to the split of the f-band of U and the relocation below the Fermi level (FL) of the hybridized states between Mo and U atoms. The characteristic distance between parallel chains, composed by consecutive U-Mo-U blocks in the the $I4/mmm$ ($2.435\mathring{A}$), changes to two distances ($2.3627\mathring{A}$ and $2.509\mathring{A}$) in the stable $Fmmm$ structure ($\delta = \pm 0.03$), (see Fig. \ref{fig:compMin}). These changes modify the U-U interaction and reproduce the same situation as that observed in $\alpha-$U where a Peierls distortion is found. S\"oderlind et al. \cite{Soderlind1995} showed that the crystal structure in pure uranium is determined by the balance between Madelung interactions and a Peierls distortion of the lattice, which favour low symmetry structures. However, whereas the Peierls distortion in $\alpha$-U is spontaneous, in U$_2$Mo it is induced by the deformation $D_6$.

\begin{figure}[!h]
\centering\includegraphics[width=5cm]{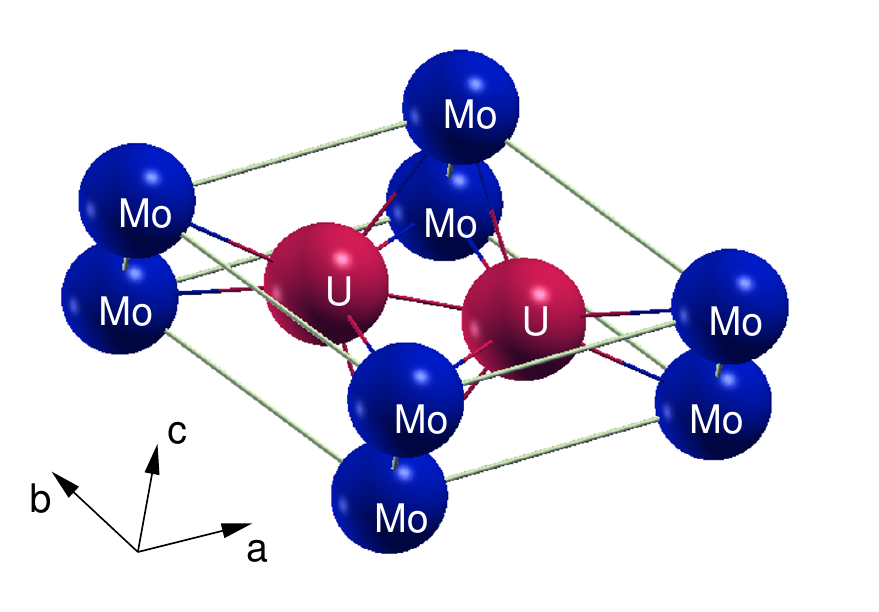}\\
\caption{$\Omega$-phase ground state structure of the U$_2$Mo compound. Space group $P6/mmm$ $(\#191)$. $a=b=4.8207\mathring{A}$ and $c=2.7643\mathring{A}$. \label{fig:Fund}}
\end{figure}

The results of this work will show that the structure of the ground state is the $P6/mmm$ space group, usually called $\Omega$-phase, depicted in Fig. \ref{fig:Fund}. 

This work presents a detailed \textit{ab initio} study of the physical properties of this novel structure in the U-Mo system, providing useful information applicable to modelling phase diagrams of multicomponent nuclear fuels. 
The stability of the $\Omega$-phase is analysed using the Phonon package and computing of the elastic constants and modules. The total energies for these calculations are obtained from the Wien2k code. The electronic properties are computed in order to explain the elastic isotropic behaviour of the new ground state structure.

This paper is organized as follows. Section \ref{sec.theory} describes the theoretical methodology. The results and discussion are presented in section \ref{sec.results}: i) Subsection \ref{subsec.ground} shows the ground state structural properties, ii) Subsection \ref{subsec.stability} discusses the elastic and phonon properties, and  iii) Subsection \ref{subsec:electronic} summarizes the electronic structure and nature of chemical bonds in the $\Omega$-phase structure. The conclusions are presented in Section \ref{sec.conclusions}.

\section{THEORETICAL METHODOLOGY} \label{sec.theory}
The total energy of U$_2$Mo in all structures, i.e. $C11b$, $Pmmn$ and $P6/mmm$, were computed using the full potential LAPW method based on the density functional theory as implemented in the WIEN2k code \cite{Wien2k}. This code uses the full-potential LAPW+lo method that makes no shape approximation to the potential or density. Electronic exchange-correlation interactions were treated within the generalized gradient approximation of Perdew, Burke and Ernzerhof \cite{Pedrew1996}, as no experimental or theoretical evidence of strong correlations was found in the system studied here. The radii of the atomic spheres (RMT) selected for U and Mo were $R_{MT} = 2.3 a.u$. and $R_{MT}  = 1.8 a.u.$, respectively. In order to describe the electronic structure of all the atoms and their orbitals, the APW + lo basis set was selected. Local orbital extensions were included to describe the semicore states. The cut-off parameter that controls convergence in the expansion of the solution to the Kohn-Sham equations was chosen to be $Rk_{max} = 8$, where $k_{max}$ is the plane wave cut-off and $R_{MT}$ is the smallest of all the atomic sphere radii. The maximum $l$ values for partial waves used inside the atomic spheres and for the non-muffin-tin matrix elements were selected to be $l_{max} = 10$ and $l_{max} = 4$, respectively. The charge density cut-off $G_{max}$ was selected as $22 Ry^{1/2}$. A mesh of $50000$ $k$-points was used in the whole Brillouin zone to study the electronic properties of the fundamental state. k-space integration was calculated using the modified tetrahedron-method \cite{Blochl1994}. The iteration process was repeated until the calculated total energy converged to less than $1\times10^{-6} Ry/cell$, and the calculated total charge converged to less than $1\times10^{-6} e/cell$. In order to calculate the internal parameters of the crystal structures, we employed the mini LAPW script implemented in the Wien2K package. The calculations were performed without including spin-orbit interactions, since no significant effects were observed near the Fermi level, as previously shown in Ref. \cite{Losada2015}.

The phonon band structure and thermal properties of the U$_2$Mo $\Omega$-phase were calculated using the finite displacement method implemented in the Phonopy \cite{phonopy} package. This method involves the creation of a supercell and introduction of atomic displacements. The 24-atom supercell was composed of $2\times 2 \times 2$ primitive unit cells. The forces acting on atoms of the supercells were calculated with the Wien2k code with a precision of $0.1 mRy/a.u.$. The phonopy post-process uses these forces to calculate phonon related properties: phonon band structure, total and partial density of states, and thermal properties.

\section{RESULTS AND DISCUSSION} \label{sec.results}

\subsection{Ground state of the U$_2$Mo compound.} \label{subsec.ground}

The instability observed in the C11b structure [3,5,6] reveals that the ground state of U$_2$Mo compound has a different crystalline structure. Wang et al., using the USPEX code, suggested the $Pmmn$ structure as the ground state . The difference in energy between the $I4/mmm$ and $Pmmn$ structures, using the Wien2k code, is 4.7 mRy. Both structures are characterized by different U-Mo distances. Whereas the distance is 3.1147$\mathring{A}$ in the $I4/mmm$ structure, the $Pmmn$ computed using the Wien2k code ($a=4.8251\mathring{A}, b=8.3504\mathring{A}c=2.7726\mathring{A} $) has two characteristic distances, one at $3.1083\mathring{A}$ and the other at 3.1118$\mathring{A}$. Although the structure proposed by Wang et al. overcame the instability, the slightly different U-Mo distances within the structure rose the question of whether another structure could be more stable than the $Pmmn$. A search was performed in the belief that the structure at $0^\circ$K should have more symmetries than the $Pmmn$. The XCrySDen \cite{xcrysden} code was used to analyse the symmetry of probable candidates, using the $Pmmn$ structure as the starting point, and computing total energy with the WIEN2k code. 
A hexagonal structure was found to be more stable than the $Pmmn$ one. It was suggested in our previous work that the $P6$ $(\#168)$ could be the ground state of the U$_2$Mo compound, as it is 0.1 mRy more stable than the $Pmmn$ structure. The $P6$ space group was used to obtain freedom in the atomic movement in order to optimize the $Pmmn$ structure. However, if all symmetries are included, the actual space group is the $P6/mmm$ $(\#191)$. This structure has the AlB$_2$ prototype usually known as $\Omega$-phase \cite{Garces2}. The structure is formed alternating close-packed hexagonal layered structures of Mo and graphite-like U layers. 

The lattice parameters of the structure are $a = b = 4.8207\mathring{A}$ and $c = 2.7643\mathring{A}$. It has one Mo atom at Wyckoff position $1(a):$ $(0, 0, 0)$ and two U atoms at positions $2(b):$ $(1/3, 2/3, z)$ and $(2/3, 1/3, z)$; where $z = 0.5$. The Mo atoms have 12 U neighbors at a distance of $3.1075\mathring{A}$. Each U atom is surrounded by 5 U atoms. Three of these are coplanar at a distance of $2.7832\mathring{A}$ and two lie along the c axis, at a distance of $2.7643\mathring{A}$. The $c$ lattice parameter used in a previous study \citep{Losada2015} ($c=2.7726\mathring{A}$) was slightly adjusted to the value presented in this work. This small difference arose 
when calculating the $C_{33}$ elastic constant with a more accurate total energy vs. $D_3$ distortion.

The structures $I4/mmm$, $Fmmm$, $Pmmn$ and $P6/mmm$ have one feature in common which is the existence of U-Mo-U blocks. Indeed, the first two structures can be interpreted as being formed by parallel linear chains along the z-directions, each one composed of consecutive U-Mo-U blocks. We found that in the $P6/mmm$ ground state structure the chains characterizing the $I4/mmm$ structure were broken and two consecutive blocks were displaced mainly perpendicular to the chains. The new structure overcame the instability by this mechanism.

\subsection{Structural stability of the U$_2$Mo $\Omega$-phase} \label{subsec.stability}
\subsubsection{Elastic properties}
Elastic constants describe the behaviour of crystals as a function of elastic deformations. Elastic properties are related to mechanical and thermal properties such as specific heat, thermal expansion, Debye temperature, melting point, and Gruneisen parameter. They also give information about the anisotropic character of bonding in crystals. 

Elastic constants are obtained by computing the second order derivatives of the total energy with respect to the distortion parameter. This calculation is very sensitive to data point selection, due to numerical inaccuracies around the minimum of total energy vs. atomic displacements. 
There are only five independent elastic constants for hexagonal crystals Ref. \cite{Nye1985}. They can be determined by imposing strains, either individually or in combination, along specific crystallographic directions. The elastic constants used in this work are $C_{11},C_{12},C_{13},C_{33}, C_{55}$, and the symmetry relationship with the remaining ones are:
\begin{align}
C_{11}&=C_{22}\\
C_{13}&=C_{23}\\
C_{55}&=C_{44}\\
C_{66}&=\frac{1}{2}(C_{11} - C_{12})\\
C_{ZZ}&=2C_{11} + 2C_{12} + 4C_{13} + C_{33}
\end{align}

The distortion matrices described by Fast et. al. \cite{Fast1995} are used in this work to compute the elastic constants for this structure. Table  \ref{tabla:matrices} summarizes the information regarding space group, distortion matrices and the third-order polynomial fitting to Wien2k results. The three distortions involving $(C_{11} + C_{22})$, $C_{33}$ and $C_{44}$ do not change the hexagonal symmetry ($P6/mmm$) but $(C_{11}-C_{12})$ and $C_{ZZ}$ do change the symmetry from hexagonal to orthorhombic and monoclinic, respectively. 
The positions of non-equivalent Mo and U atoms in the hexagonal $P6/mmm$ space group are fixed in special positions, and consequently, relaxation of internal coordinates is not necessary. The Mo atom positions remain fixed by symmetry to the corner positions of the crystal cell for distortions $C_{11}-C_{12}$ and $C_{ZZ}$, but the position of U atoms should be fully relaxed in order to compute these elastic constants. The relaxation 
procedure, as implemented in the code Wien2k, was stopped for interatomic forces lower than $2 mRy/a.u$.  Table \ref{tabla:ElastConst} compares the results of this work for the $P6/mmm$ structure with the $Pmmn$ one \citep{Wang2014}. 

\begin{table*}[tbp]
\centering
\begin{tabular}{
 | >{\centering\arraybackslash}m{2cm} |
   >{\centering\arraybackslash}m{5cm} |
   >{\centering\arraybackslash}m{2cm} |
   >{\centering\arraybackslash}m{5.5cm} |
}
\hline
Combination of elastic constants & Deformation matrix & Space group & Polynomial fit \\
\hline
 $C_{11}+C_{12}$ & \begin{equation}
	D_1\,=\,
	\left(
	 \begin{array}{ccc}
		{1+\delta} & 0 & 0\\
		0 & {1+\delta} & 0\\
		0 & 0 & 1\\
     \end{array}
    \right) \; \nonumber
\end{equation} & $P6/mmm$ $(\#191)$ & {\centering\includegraphics[scale=0.4]{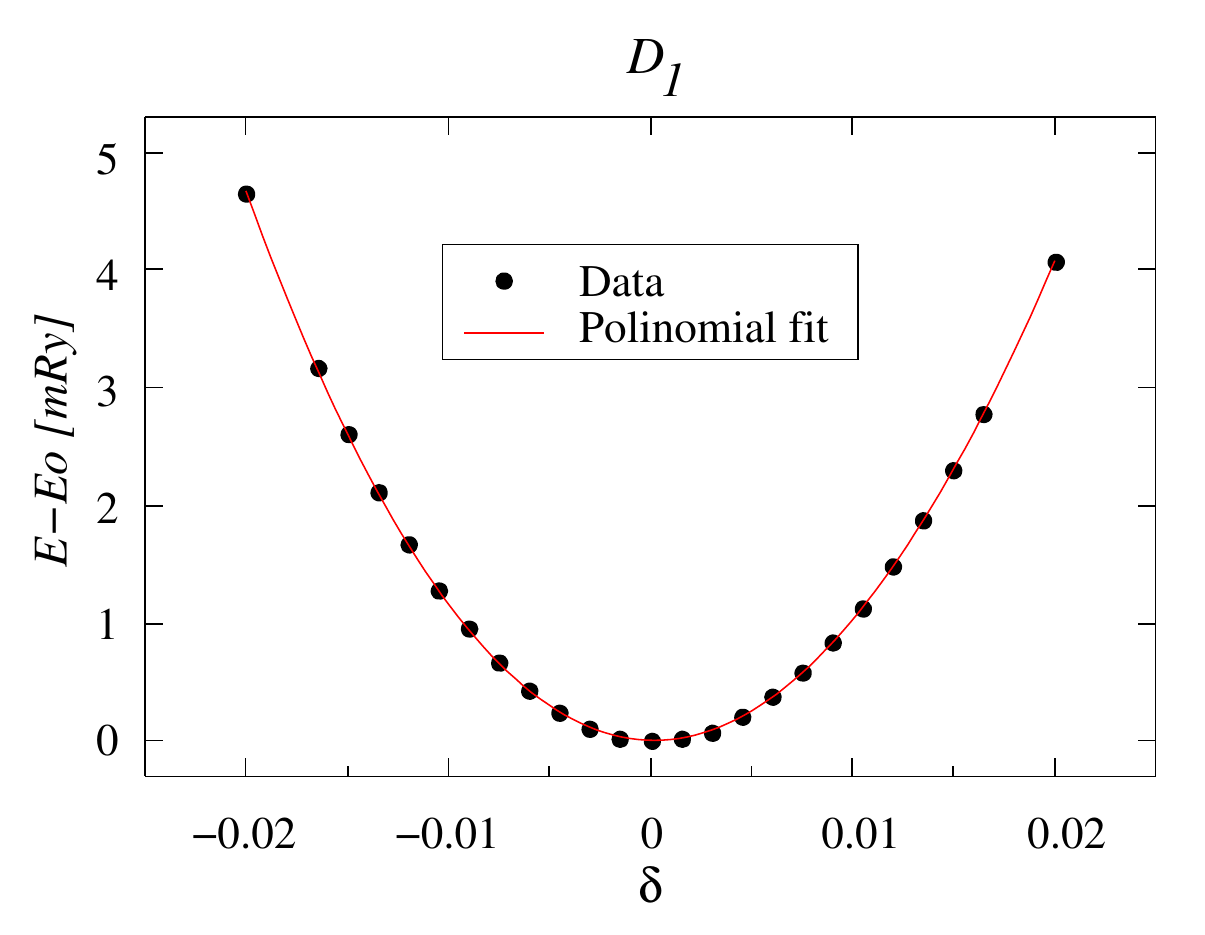}}  \\
\hline
 $C_{11}-C_{12}$ & \begin{equation}
	D_2\,=\,
	\left(
	 \begin{array}{ccc}
		{1+\delta} & 0 & 0\\
		0 & {1-\delta} & 0\\
		0 & 0 & 1\\
     \end{array}
    \right) \; \nonumber
\end{equation} & $Cmmm$ $(\#65)$ & {\centering\includegraphics[scale=0.4]{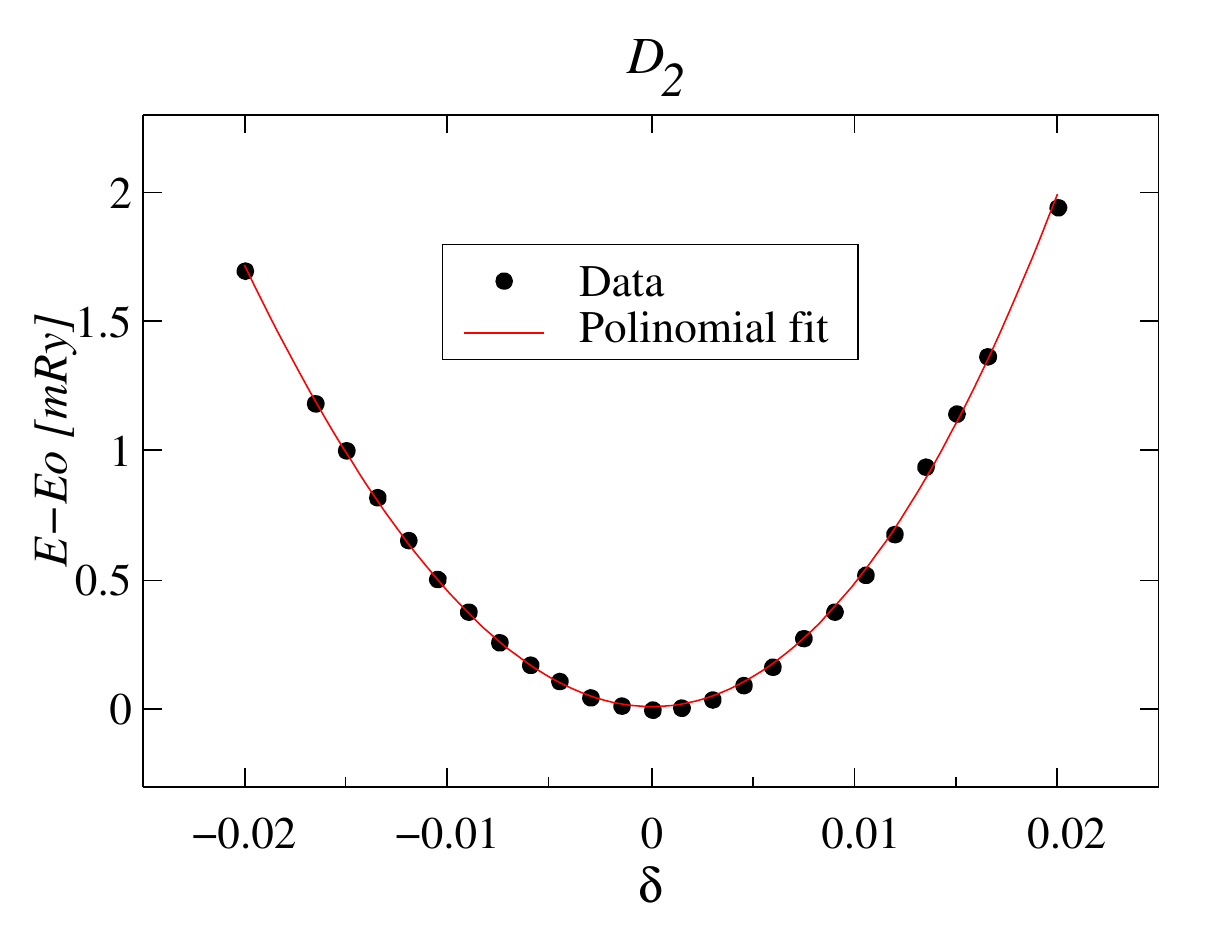}}  \\
\hline
 $C_{33}$ & \begin{equation}
	D_3\,=\,
	\left(
	 \begin{array}{ccc}
		1 & 0 & 0\\
		0 & 1 & 0\\
		0 & 0 & {1+\delta}\\
     \end{array}
    \right) \; \nonumber
\end{equation} & $P6/mmm$ $(\#191)$ & {\centering\includegraphics[scale=0.4]{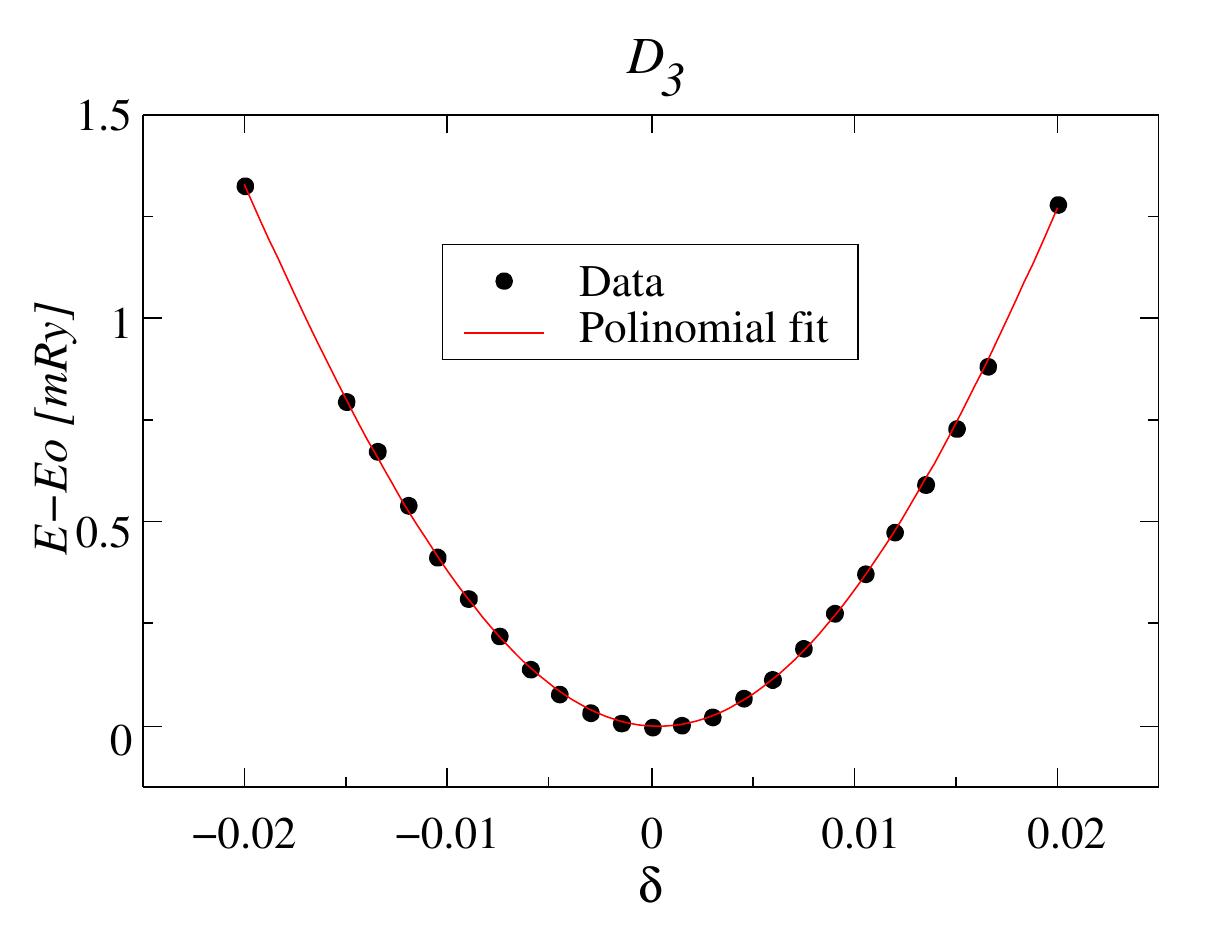}}  \\
\hline
 $C_{55}$ & \begin{equation}
	D_4\,=\,
	\left(
	 \begin{array}{ccc}
		1 & 0 & \delta\\
		0 & 1 & 0\\
		\delta & 0 & 1\\
     \end{array}
    \right) \; \nonumber
\end{equation} & $C2/m$ $(\#12)$ & {\centering\includegraphics[scale=0.4]{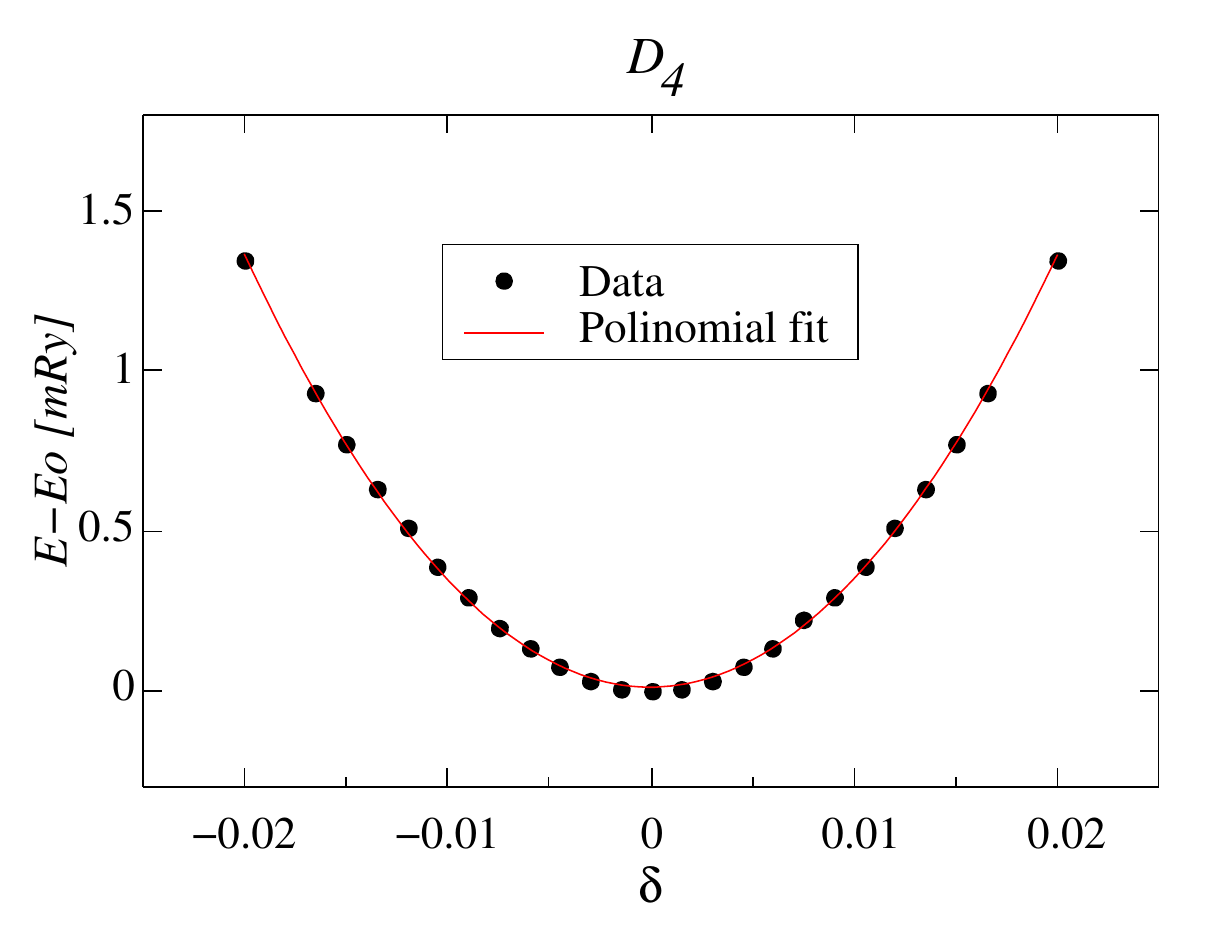}}  \\
\hline
 $C_{ZZ}$ & \begin{equation}
	D_5\,=\,
	\left(
	 \begin{array}{ccc}
		{1+\delta} & 0 & 0\\
		0 & {1+\delta} & 0\\
		0 & 0 & {1+\delta}\\
     \end{array}
    \right) \; \nonumber
\end{equation} & $P6/mmm$ $(\#191)$ & {\centering\includegraphics[scale=0.4]{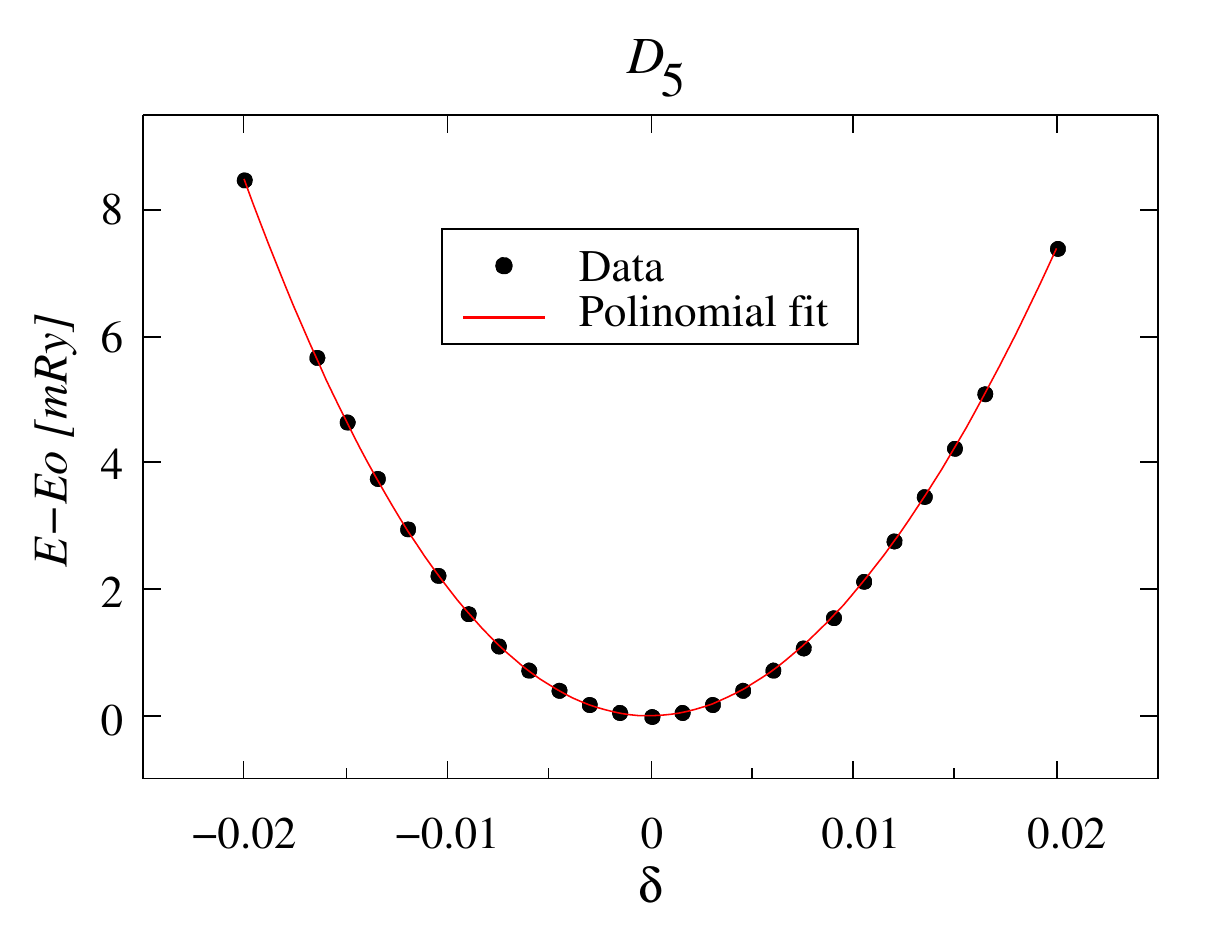}}  \\
\hline
\end{tabular}
\captionsetup{justification=centering,margin=1cm}
\caption{Distortion matrices, space group, Wien2k results and third order polynomial fit used to compute the five elastic constants of the $P6/mmm$ ground state. $C_{ZZ}$ stands for the following elastic constants combination: $C_{ZZ}=2C_{11}+2C_{12}+4C_{13}+C_{33}$.} \label{tabla:matrices}
\end{table*} 

The energy associated with the five distortions and the elastic constants deduced from them are summarized below. In the following equations,  $V_0$ is the volume of the unstrained system, $E(V_0 , \delta )$ is the total energy computed by the ab initio code, and $\tau_1$, $\tau_2$, $\tau_3$ and $\tau_5$ are the elements of the stress tensor in Voigt notation ($\tau_1=\tau_{xx}$, $\tau_2=\tau_{yy}$, $\tau_3=\tau_{zz}$ and $\tau_5=\tau_{xz}$).

\begin{equation}
E(V,\delta) = E(V_0,0) + V_0\left[ (\tau_1 + \tau_2)\,\delta + (C_{11} + C_{12})\,\delta^2\right]
\end{equation}

\begin{equation}
C_{11} + C_{12} = \frac{1}{2\,V_0}\,\frac{d^2(E-E_0)}{d\delta^2}
\end{equation}

\begin{equation}
E(V,\delta) = E(V_0,0) + V_0\left[ (\tau_1 - \tau_2)\,\delta + (C_{11} - C_{12})\,\delta^2\right]
\end{equation}

\begin{equation}
C_{11} - C_{12} = \frac{1}{2\,V_0}\,\frac{d^2(E-E_0)}{d\delta^2}
\end{equation}

\begin{equation}
E(V,\delta) = E(V_0,0) + V_0\left[ \tau_3\,\delta + \frac{C_{33}}{2}\,\delta^2\right]
\end{equation}

\begin{equation}
C_{33} = \frac{1}{V_0}\,\frac{d^2(E-E_0)}{d\delta^2}
\end{equation}

\begin{equation}
E(V,\delta) = E(V_0,0) + V_0\left[ \tau_5\,\delta + 2\,C_{55}\,\delta^2\right]
\end{equation}

\begin{equation}
C_{55} = \frac{1}{4\,V_0}\,\frac{d^2(E-E_0)}{d\delta^2}
\end{equation}

\begin{equation}
E(V,\delta) = E(V_0,0) + V_0\left[ (\tau_1 + \tau_2 + \tau_3)\,\delta + \frac{1}{2}\,C_{ZZ}\,\delta^2\right]\,
\end{equation}

\begin{equation}
C_{ZZ} = \frac{1}{V_0}\,\frac{d^2(E-E_0)}{d\delta^2}
\end{equation}
\\

The deformation energy $\left(\frac{1}{2}C_{ij}\,\,\epsilon_i\,\epsilon_j\right)$ must be positive definite to ensure the mechanical stability of the crystal. This implies further restrictions on the $C_{ij}$ found by algebraical methods. For a crystal with hexagonal symmetry, the constrains \cite{Nye1985} are: \\
\begin{equation}
C_{44} > 0,\;\;  C_{11} > \vert C_{12} \vert,\;\;\; (C_{11} + C_{12})C_{33} > 2C_{13}^2
\end{equation}

\begin{table*}[!h]
\centering
\begin{tabular}{
 | >{\centering\arraybackslash}m{1.6cm} |
   >{\centering\arraybackslash}m{1cm} |
   >{\centering\arraybackslash}m{1cm} |
   >{\centering\arraybackslash}m{1cm} |
   >{\centering\arraybackslash}m{1cm} |
   >{\centering\arraybackslash}m{1cm} |
   >{\centering\arraybackslash}m{1cm} |
   >{\centering\arraybackslash}m{1cm} |
   >{\centering\arraybackslash}m{1cm} |
   >{\centering\arraybackslash}m{2cm} |
}
\hline
 Space group & $C_{11}$ & $C_{12}$ & $C_{13}$ & $C_{22}$ & $C_{23}$ & $C_{33}$ & $C_{44}$ & $C_{66}$  & Reference \\
\hline
 $P_6/mmm$ & $304$ & $123$ & $103$ & $304^*$ & $103^*$ & $288$ & $66$ & $90^*$ & This work.  \\
 \hline
 $Pmmn$ & $299$ & $131$ & $116$ & $293$ & $17$ & $246$ & $74$ & $87$ & Wang et. al.\citep{Wang2014} \\
\hline
\end{tabular}
\captionsetup{justification=centering,margin=1cm}
\caption{Comparison between the elastic constants calculated in this work for the $P6/mmm$ structure  
and the $Pmmn$ one proposed by Wang et al. \\($^*$ stands for constants obtained by symmetry operations.)  } \label{tabla:ElastConst}
\end{table*}

U$_2$Mo in the $P6/mmm$ structure is mechanically stable, since all its elastic constants fulfil the above requirements. The elastic constants $C_{11}$ and $C_{33}$ have the largest values and represent the elasticity 
along directions parallel and perpendicular to the basal plane; $x-$ and $z-$  directions respectively. The result can be interpreted as showing that the bonding characters along the x- and z-directions are the two strongest of all the directions. The other three elastic constants, $C_{12}$, $C_{13}$ and $C_{44}$, are related to the elasticity of shape deformation. Therefore, their similar values show that the crystal presents almost the same response to this kind of deformation.

Based on the calculated elastic constants, the theoretical polycrystalline bulk modulus $B$ and shear modulus $G$ can be determined from their sets of elastic constants using the approximations of Voigt \citep{Voigt1928}, Reuss \citep{Reuss1929} and Hill \citep{Hill1952}. While the Voigt approach gives the upper bound of elastic properties,  the Reuss approach provides the lower bound.  
For the Voigt average, the bulk and shear moduli for hexagonal systems are given by:

\begin{equation}
B_V = \frac{1}{9}\left[ 2(C_{11}+C_{12})+C_{33}+4C_{13} \right]
\end{equation}
\begin{equation}
G_V = \frac{1}{30}\left( M + 12C_{44} +12C_{66} \right)
\end{equation}
For the Reuss average \citep{Reuss1929}, the moduli are given by,
\begin{equation}
B_R = \frac{C^2}{M}
\end{equation}
\begin{equation}
G_R = \frac{5}{2}\frac{C^2\,C_{44}\,C_{66}}{3\,B_V\,C_{44}\,C_{66} + C^2\,(C_{44}+C_{66})}
\end{equation}

\begin{equation}
M = C_{11} + C_{12} + 2\,C_{33} - 4\,C_{13}
\end{equation}
\begin{equation}
C^2 = (C_{11} + C_{12})\,C_{33} - 2\,C_{13}^2
\end{equation}

Hill proved that the Voigt and Reuss methods resulted in the theoretical upper and lower bounds of the isotropic elastic modulus, respectively. Hill suggested a practical estimation of the bulk and shear moduli as the arithmetic means of both values. The bulk and shear moduli are computed in the Hill empirical average as,
\begin{equation}
B_H = (B_V + B_R)/2 \,,
\end{equation}
\begin{equation}
G_H = (G_V + G_R)/2 \, .
\end{equation}
From these values, Young's modulus $E$ and Poisson's ratio $\nu$ can be computed as, 
\begin{equation}
E = \frac{9\,B_H\,G_H}{3\,B_H + G_H}
\end{equation}
\begin{equation}
\nu = \frac{3\,B_H - 2\,G_H}{2\,(3\,B_H + G_H)}
\end{equation}

Based on the fact that the shear modulus $G$ represents the resistance to plastic deformation while the bulk modulus represents resistance to fracture, Pugh \citep{Pugh1954} introduced the ratio between the bulk and shear moduli, $B/G$, for polycrystalline phases as a measure of the fracture/toughness of metals. A value greater that $1.75$ is associated with ductility and a lower brittleness behaviour. This value for U$_2$Mo in the $P6/mmm$ structure is $B/G=2.13$.
This result indicates that U$_2$Mo has ductile behaviour. The ductility of this compound is confirmed by the values of Poisson's ratio, a property also used to measure the ductility and brittleness of materials \citep{Gao2014}. The critical value which separates ductility from brittleness is $0.26$ \citep{Hao2007}. The value of $\nu = 0.3$, showed in Table \ref{tabla:ElastConst2}, confirms the ductile character of the U$_2$Mo compound with the $\Omega$-phase structure.

Analysis of elastic anisotropy allows understanding of the bonding nature of a crystal. It can be described either through the universal anisotropic index $A^U$ proposed by Ranganathan and Ostioja-Starzewski \citep{Ranganathan2008} for crystals with any symmetry defined by,
\begin{equation}
A^U = 5\,\frac{G_V}{G_R}+\frac{B_V}{B_R}-6 \geqslant 0\,,
\end{equation}
or through the percent anisotropy indexes of bulk and shear moduli ($A_B$ and $A_G$), proposed by Chung and Buessen \citep{Chung1967,Chung2}, defined by,
\begin{align}
A_B &= \frac{B_V - B_R}{B_V + B_R}\,,\\
A_G &= \frac{G_V - G_R}{G_V + G_R}\,.
\end{align}

\begin{table*}[!h]
\centering
\begin{tabular}{
 | >{\centering\arraybackslash}m{1.6cm} |
   >{\centering\arraybackslash}m{1cm} |
   >{\centering\arraybackslash}m{1cm} |
   >{\centering\arraybackslash}m{1cm} |
   >{\centering\arraybackslash}m{1cm} |
   >{\centering\arraybackslash}m{1cm} |
   >{\centering\arraybackslash}m{1cm} |
   >{\centering\arraybackslash}m{1cm} |
   >{\centering\arraybackslash}m{1cm} |
}
\hline
 Space group & $B$ & $G$ & $E$ & $\nu$ & $B/G$ & $A^U$ & $A_B$ & $A_G$  \\
\hline
 $P_6/mmm$ & $173$ & $81$ & $210$ & $0.3$ & $2.13$ & $0.15$ & $.00139$ & $0.0144$  \\
 \hline
\end{tabular}
\captionsetup{justification=centering,margin=1cm}
\caption{Elastic moduli and anisotropy indices for the $\Omega$-phase of U$_2$Mo compound} \label{tabla:ElastConst2}
\end{table*}

The Voigt and Reuss approximations should give the same values for $B$ and $G$ modules for isotropic structures.  Consequently, deviations from zero indicate anisotropy. The results for the anisotropic indexes are shown in Table \ref{tabla:ElastConst2}. It can be seen that $A_U=0.15$ and the percent anisotropy, both in shear $A_G= 0.014$ and compression $A_B=0.0014$, are almost zero suggesting that U$_2$Mo in the $P6/mmm$ structure is an elastic isotropic material. The minimal degree of anisotropy and the values of elastic constants exhibited can be understood by examining the nature of chemical bonds between atoms.

\subsubsection{Phonon dispersion relation}

\begin{figure*}[!h]
\centering
\parbox{4cm}{\centering
\includegraphics[scale=0.3]{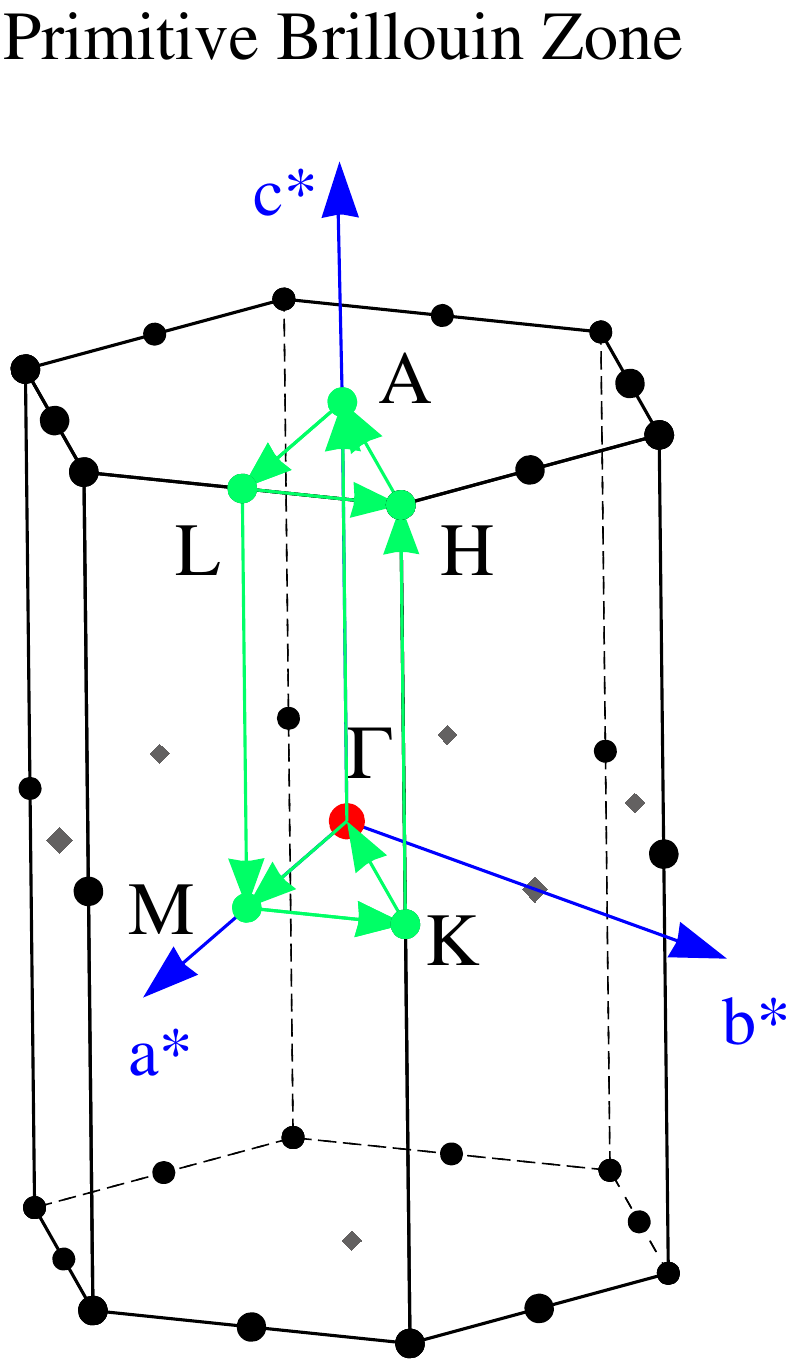}
\captionsetup{justification=centering}
\caption{Brillouin zone of hexagonal type lattice \cite{Setyawan2010299} with the selected path for phonon dispersion curves and band structure calculations.}
\label{fig:totalpath}}
\begin{minipage}{12cm}
\includegraphics[width=12cm]{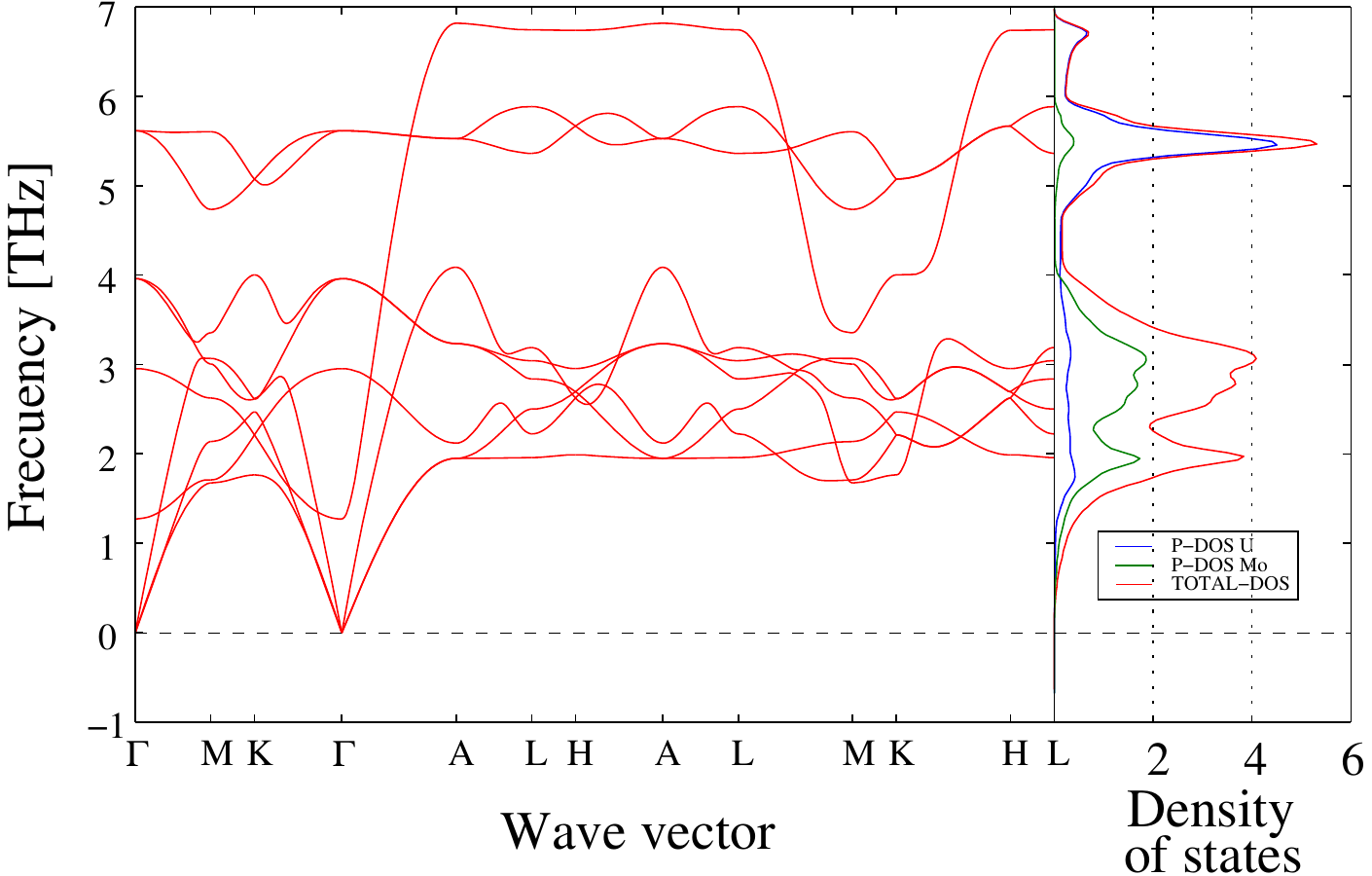}
\captionsetup{justification=centering}
\caption{Calculated phonon dispersion curves (left) and projected phonon density of states (right) of  the U$_2$Mo $\Omega$-phase.}
\label{fig:Phonons}
\end{minipage}
\end{figure*}

The dynamical stability of U$_2$Mo in the $P6/mmm$ structure requires the energies of phonons to be positive for all wave vectors in the Brillouin zone (BZ). The phonon dispersion relations are computed with the Phonopy code \cite{phonopy} by constructing a $2\times 2\times 2$ supercell. The path in the reciprocal lattice along the high-symmetry points of the hexagonal BZ considered to calculate the phonon dispersion is $\Gamma - A - L - H - A - L - M - K - H - L,$ and is depicted in figure \ref{fig:totalpath}. The full phonon dispersion of the U$_2$Mo compound consists of $9$ branches: $3$ acoustic and $6$ optics. The results obtained are shown in Fig. \ref{fig:Phonons}.
The phonon dispersion curves do not present imaginary frequencies, showing the dynamical stability of the 
$\Omega$-phase of the U$_2$Mo compound, in agreement with the results obtained by analysing elastic constants. 
The phonon densities of state can be clearly separated by Mo and U due to the very different masses of U and Mo atoms. The low-frequency region (below $4 THz$) is dominated by the U atom because the acoustic modes originate mainly from heavy elements, and above $4 THz$ the phonon modes are mainly from the Mo atoms, in agreement with previous results \citep{Wang2014}.

\subsection{Electronic structure and nature of chemical bonds.}\label{subsec:electronic}
The results of previous sections present the $P6/mmm$ structure of the U$_2$Mo compound as a ductile material with a minimum degree of anisotropy. The reasons behind these properties can be understood by examining the electronic properties and chemical bonding between the U and Mo atoms.

\begin{figure}[h!]
\centering
\parbox{8cm}{\centering
\includegraphics[width=8cm]{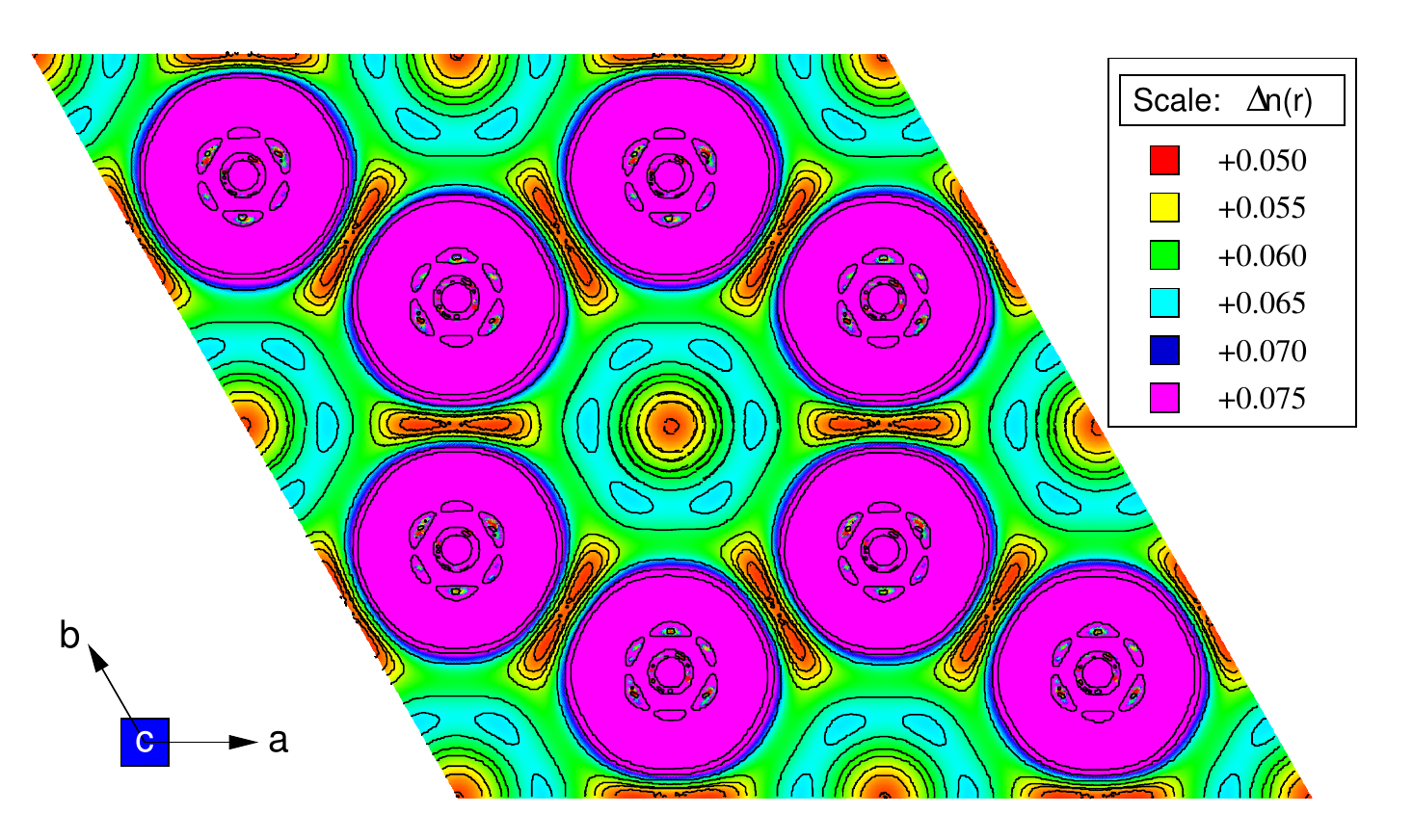}
\captionsetup{justification=centering}
\caption{Charge density distribution on the (0 0 0 $2$) plane of the $\Omega$-phase.}
\label{fig:dens} }
\qquad
\begin{minipage}{8cm}
\centering
\includegraphics[width=8cm]{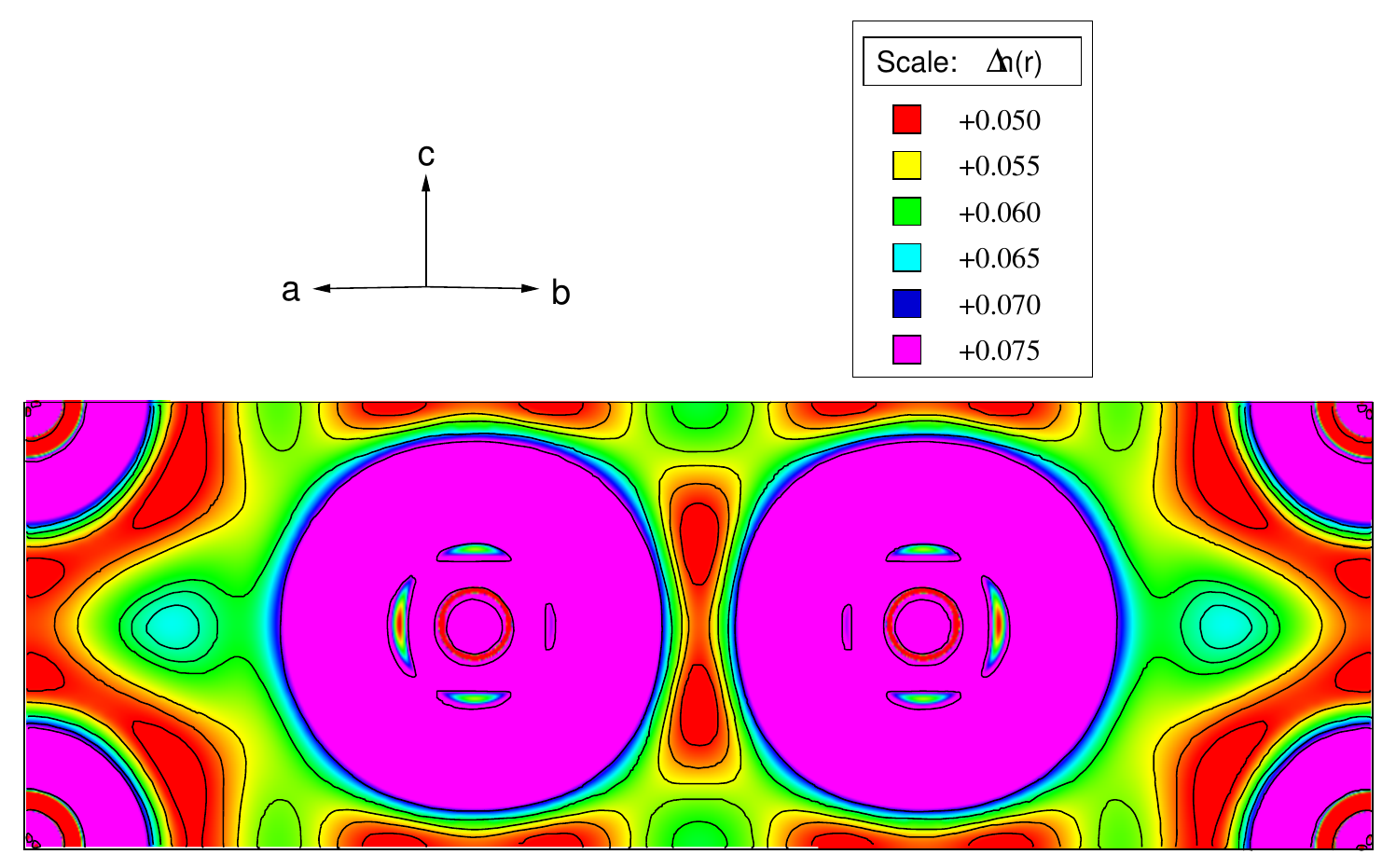}
\captionsetup{justification=centering}
\caption{Charge density distribution on the (1 1 $\bar{2}$ 0) plane of the $\Omega$-phase.}
\label{fig:dens2}
\end{minipage}
\end{figure}

Figures \ref{fig:dens} and \ref{fig:dens2} show the charge density plots for planes with Miller-Bravais indices (0 0 0 $2$) and (1 1 $\bar{2}$ 0). The charge density is strongly directional and concentrated between U and Mo planes with chemical bonds of lower strength between U-U and Mo-Mo atoms. The charge density distribution between U atoms in the (0 0 0 $2$) plane and along the c-axis are similar, as shown in Figs. \ref{fig:dens} and \ref{fig:dens2}. The result of the strong bond between U and Mo is such that the structure is characterized by one Mo atom surrounded by twelve U, each U being surrounded by six Mo atoms and five U atoms. The charge density plots reveal the main reason for the elastic isotropic behaviour of this structure: the U and Mo atoms have an almost chemically isotropic environment.

\begin{figure}[!h]
\centering
\includegraphics[width=8cm]{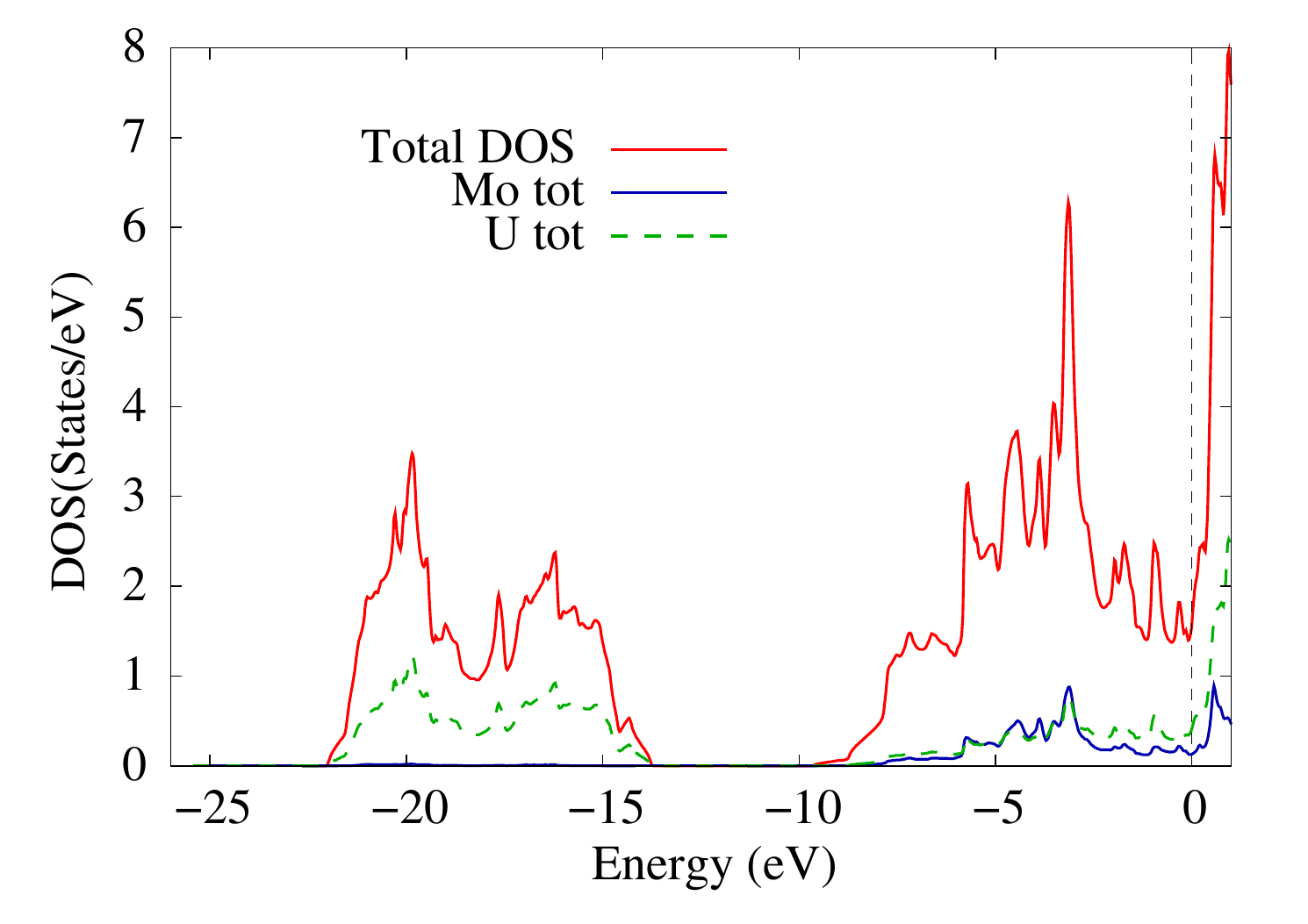}
\vspace{-0.3cm}
\captionsetup{justification=centering}
\caption{Calculated total density of states (red line) and atom-projected densities (Mo: blue line, U: green line) of electron states for the $\Omega$-phase of the U$_2$Mo compound.}
\label{fig:TotalDdos} 
\end{figure}

The strength of the interatomic bonds and elastic properties of materials are determined mainly by the occupied states at and near the Fermi level. Consequently, analysis of the DOS below the Fermi level will identify the electronic state responsible for the symmetric charge density distribution and the consequent elastic isotropic behaviour of U$_2$Mo.

\begin{figure}[!h]	
\captionsetup{justification=centering}
	\centering
	\vspace{0.3cm}
	\begin{subfigure}[t]{8cm}
		\centering
		\includegraphics[scale=0.8]{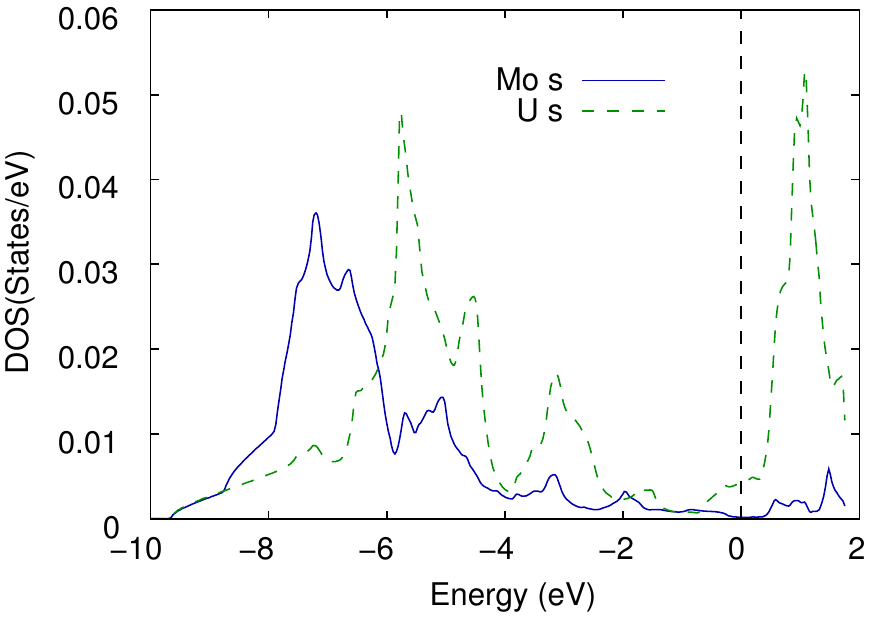}
		\caption{s states}\label{fig:sdos}	
	\end{subfigure}
    \quad 
	\begin{subfigure}[t]{8cm}
		\centering
		\includegraphics[scale=0.8]{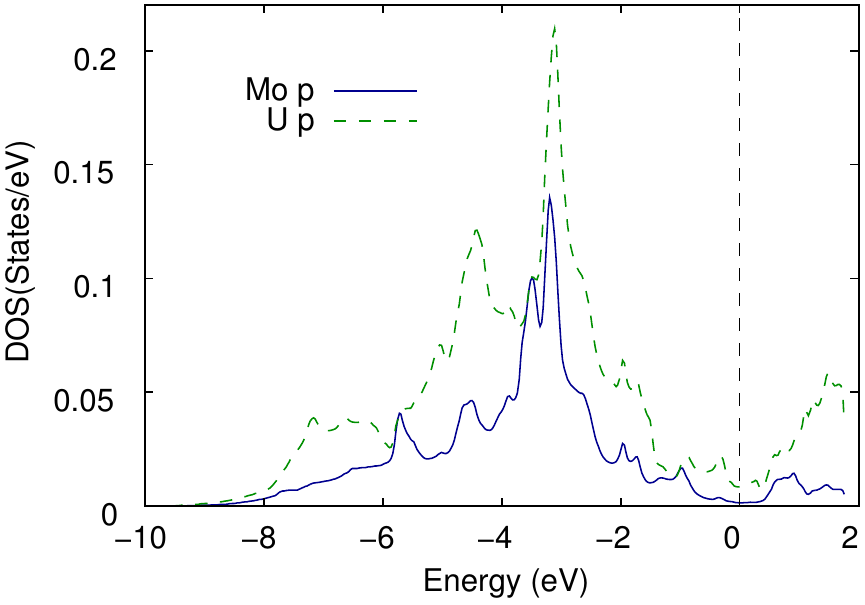}
		\caption{p states}\label{fig:pdos}	
	\end{subfigure}
	\quad \\
	\begin{subfigure}[t]{8cm}
		\centering
		\includegraphics[scale=0.8]{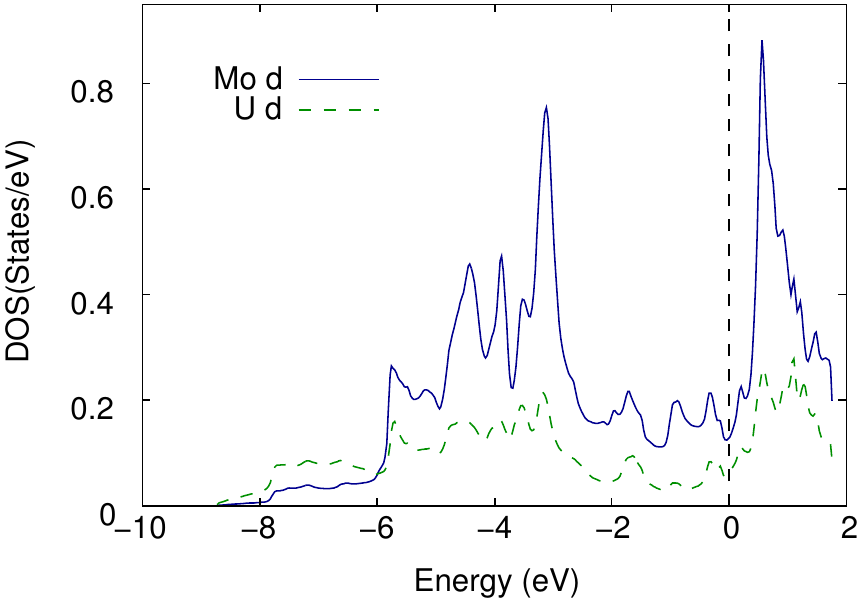}
    	\caption{d states}\label{fig:ddos}	
	\end{subfigure}
	\quad
	\begin{subfigure}[t]{8cm}
		\centering
		\includegraphics[scale=0.8]{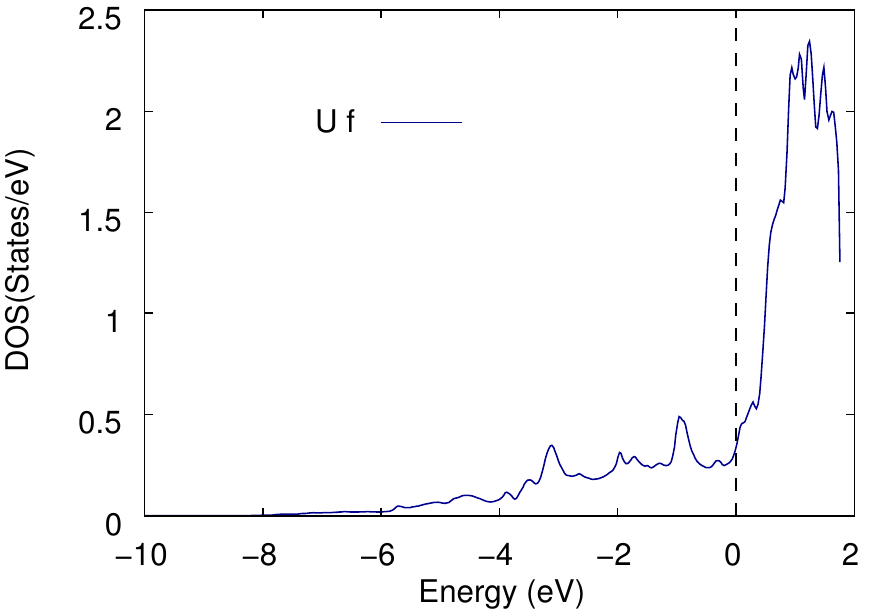}
		\caption{f states}\label{fig:fdos}	
	\end{subfigure}
	\caption{Projected densities of states for Mo and U atoms in the range $-10eV$ to $2 eV$: a) s-states, b) p-states, c) d states and d) f-states. }\label{fig:projdos}
\end{figure}

Figs. \ref{fig:TotalDdos} and \ref{fig:projdos} show the total and projected DOS for the different electronic states of Mo and U, respectively. The main contributions to the DOS near the Fermi level are due to the d-orbitals of Molybdenum and the d- and f-orbitals of Uranium. The projected DOS show that the location of these orbitals could be hybridized, as they are located almost in the same energy range near the FL. The band character will be computed along the path showed in Fig. \ref{fig:totalpath}, in order to study this possibility. The remaining electronic states have negligible contributions near the FL. However, the s- and p- electronic states of both atoms make their main contribution between $4-5eV$ below the FL.

\begin{figure}[!h]
\centering
\parbox{8cm}{\centering
\includegraphics[width=8cm]{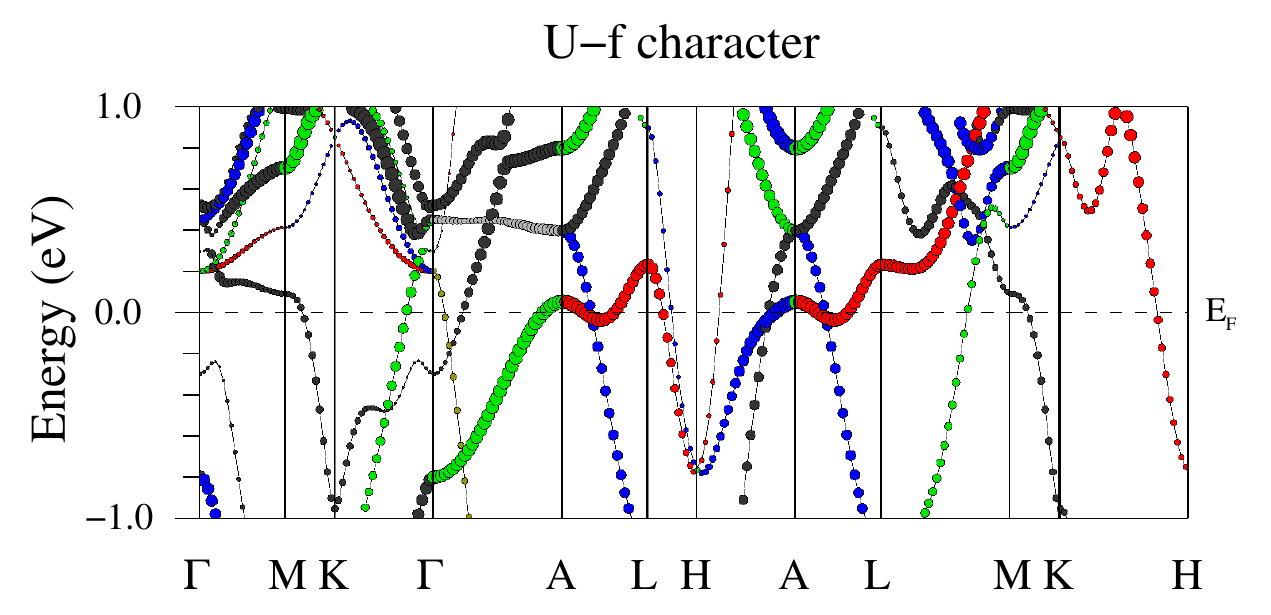}\\
\includegraphics[width=8cm]{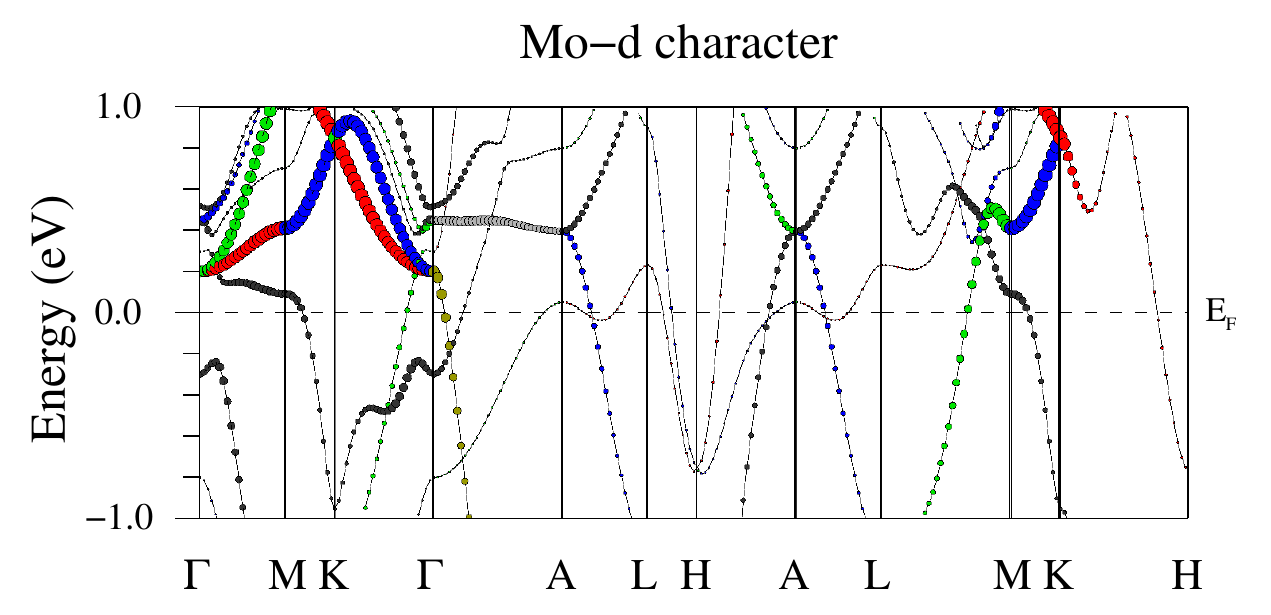}\\ 
\includegraphics[width=8cm]{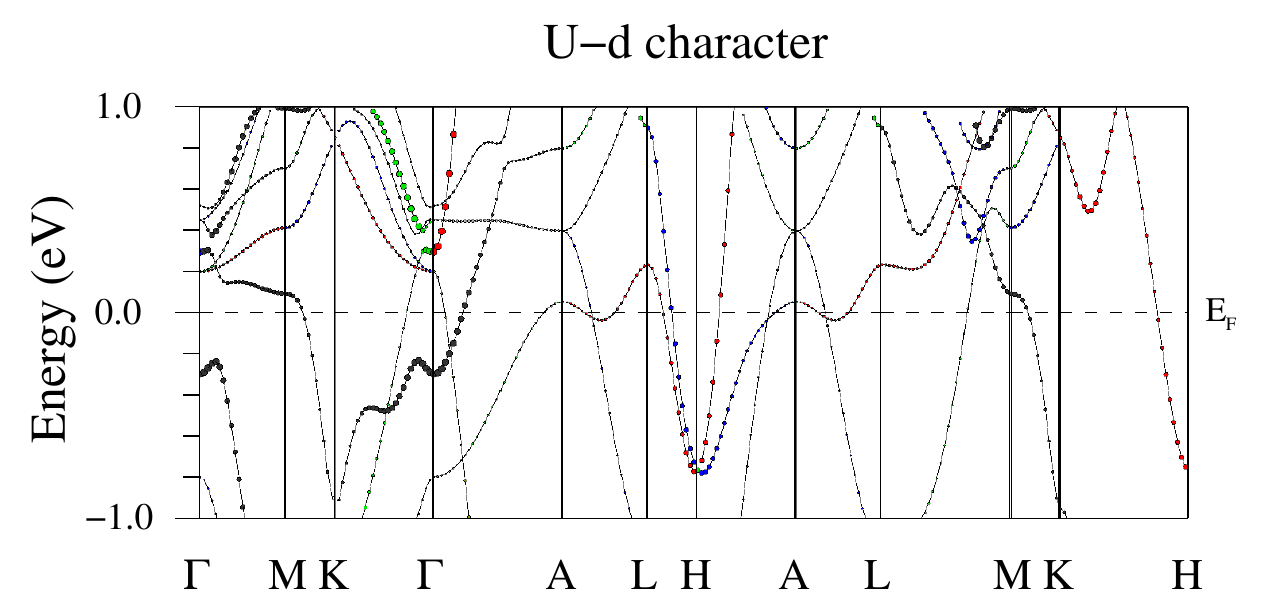}\\
}
\captionsetup{justification=centering}
\caption{Predominant character of the band structure near the Fermi level obtained following the path depicted in Fig. \ref{fig:totalpath}. Number of $k$-points $=50000$. a) f character of U, b) d character of Mo  and c) d character of U.}
\label{fig:bands}
\end{figure}

Fig. \ref{fig:bands} shows the character of the bands for the relevant states near the FL. Whereas the only relevant orbital for Mo is the d-state, the main contribution for U comes from both the d- and f-states. Figure \ref{fig:bands} shows strong hybridization between f-U and d-Mo states along the line $\Gamma-M-K-\Gamma$, and also along the $A-L$, $L-M$ and $M-K$ lines. Hybridized states between d-Mo and d-U can be observed mainly along the $K-\Gamma$ direction.

In summary, the bands with character plot show significant hybridization between d-Mo and d- and f-U states. Although the ground state structure is characterized by strong bonding and almost isotropic U-Mo interactions, the $P6/mmm$ structure becomes unstable by increasing the temperature stabilizing the $I4/mmm$ one.

\section{CONCLUSIONS} \label{sec.conclusions}
The U-Mo system has technological relevance as a potential fuel in experimental reactors and in GenIV reactors. The unexpected physical behaviour found recently in the ground state reveals that the system is still not fully characterized. The available information suggests the $C11b$ (MoSi$_2$ prototype $I4/mmm$ space group) structure as the ground state of this compound. It was shown, however, in Refs. \citep{Losada2015,Wang2014,Liu2015} that the assumed $C11b$ structure is unstable under the $D_6$ distortion associated with the $C_{66}$ elastic constant. The root causes of this structural instability are related to the existence of a Peierls distortion induced by the deformation $D_6$ \citep{Losada2015}. The results of this work show the $P6/mmm$ space group, usually called $\Omega$-phase, to be the structure of the ground state.
This work presents a detailed \textit{ab initio} study of the U-Mo ground state, in order to identify its main physical properties. This information can be useful in modelling the U-Mo system and multicomponent phase diagram of nuclear fuels, mainly with theoretical tools based on ground state properties.

The stability of the $\Omega$-phase is analysed by computing the elastic constants. All of these verify the stability criterion. Analysis of the isotropic indexes shows that the new structure is a ductile material which has a minimal degree of anisotropy, suggesting that U$_2$Mo in the $P6/mmm$ structure is an elastic isotropic material. The phonon dispersion spectrum does not show imaginary frequencies, supporting the dynamic stability found by analysing the elastic constants. The phonon densities of state are clearly separated by Mo and U due to the very different masses of U and Mo atoms. The low-frequency region (below $4THz$) is dominated by the U atom because the acoustic modes originate mainly from heavy elements, while above $4THz$ the phonon modes are mainly from the Mo atoms, in agreement with previous results \citep{Wang2014}.
The electronic properties such as DOS, charge density and character of bands reveal a high level of hybridization between the d-electronic states of Molybdenum and the d- and f-states of Uranium atoms. The charge density is strongly directional and concentrated between U and Mo planes, with lower-strength chemical bonds between U-U and Mo-Mo atoms. The charge density plots reveal the main reason for the elastic isotropic behaviour of this structure, i.e. the U and Mo atoms have an almost chemically isotropic environment.

The unexpected instability of the $C11b$ compound and the existence of a new ground state with the $\Omega$-phase structure raises the question of whether other compounds with different compositions and structures may exist in the ground state of the U-Mo system. Furthermore, detailed research is needed to fully characterize the ground state of this system and to understand the relationship between the ground state properties, the superconductivity of metastable phases \cite{Berlincourt195912,Adamska2014} and the asymmetric shape of the heat of formation vs. Mo concentration for the disordered bcc solid solution \cite{Landa20111}.

\begin{multicols}{2}
\label{sec.references}
\bibliographystyle{apsrev4-1}
\bibliography{bibliografia.bib}
\end{multicols}

\end{document}